\documentclass{optica-article}

\journal{opticajournal} 

\articletype{Research Article}

\usepackage{multirow}
\usepackage{lineno}
\usepackage[normalem]{ulem}

\begin{document}

\title{Silicon-based vacuum window for millimeter and submillimeter-wave astrophysics}


\author{
Ryota Takaku\authormark{1}, 
Scott Cray\authormark{2},
Kosuke Aizawa\authormark{3},
Akira Endo\authormark{4},
Shaul Hanany,\authormark{2},
Kenichi Karatsu\authormark{5},
J\"{u}rgen Koch,\authormark{6},
Kuniaki Konishi\authormark{7}, 
Tomotake Matsumura\authormark{8,9,10}, 
and Haruyuki Sakurai\authormark{7}
}

\address{
\authormark{1}Okayama University, 3-1-1 Tsushima-naka, Kita-ku, Okayama 700-8530, Japan

\authormark{2}School of Physics and Astronomy, University of Minnesota, Twin Cities, 115 Union St. SE, Minneapolis MN 55455, USA\\

\authormark{3}Department of Physics, The University of Tokyo, 7-3-1 Hongo, Bunkyo-ku, Tokyo 113-8654, Japan\\

\authormark{4}Department of Microelectronics, Faculty of Electrical Engineering, Mathematics and Computer Science, Delft University of Technology,HB.18.270, Mekelweg 4, 2628 CD Delft, The Netherlands\\
\authormark{5}Netherlands Institute for Space Research (SRON), Niels Bohrweg 4 2333 CA Leiden, Netherlands\\

\authormark{6}Laser Zentrum Hannover, Hollerithallee 8 D-30419, Hannover, Germany\\

\authormark{7}Institute for Photon Science and Technology (IPST), The University of Tokyo, 7-3-1 Hongo, Bunkyo-ku, Tokyo 113-0033, Japan\\

\authormark{8} Kavli Institute for the Physics and Mathematics of the Universe (WPI), The University of Tokyo, 5-1-5 Kashiwa-no-Ha, Kashiwa, Chiba 277-8583, Japan\\
\authormark{9}Center for Data Driven Discovery (CD3), Kavli Institute for the Physics and Mathematics of the Universe (IPMU), The University of Tokyo, 5-1-5 Kashiwa-no-Ha, Kashiwa, Chiba 277-8583, Japan\\
\authormark{10}ILANCE, CNRS, University of Tokyo International Research Laboratory, Kashiwa, Chiba 277-8582, Japan\\

}
\email{\authormark{*}ryota.takaku@okayama-u.ac.jp} 

\begin{abstract*} 

We designed, fabricated, and characterized the properties of a silicon-based vacuum window suitable for millimeter-wave astrophysical applications. The window, which has a diameter of 124~mm, optically active diameter of 68~mm, and thickness of about 4~mm, gives an average transmittance and reflectance of 99\% and 1\%, respectively, a fractional bandwidth of 67\%. Absorptive loss is below the detection limit of our measurement. The anti-reflection coating is made with laser ablated sub-wavelength structures (SWS), and the measured transmittance and reflectance values agree with modeling based on the measured SWS shapes.  
The window has been integrated into DESHIMA v2.0, an  astrophysics instrument that took year-long observations with the Atacama Submillimeter Telescope Experiment. 

\end{abstract*}


\section{INTRODUCTION}
\label{sec:intro}  
It is common for millimeter-wave astrophysical instruments to have cryogenic dewars with transparent vacuum windows~\cite{Hanany2013,kerr1992_edir292,Runyan_2003}.
Ideally, the vacuum window should be mechanically robust to withstand the atmospheric differential pressure, and have high transmittance. 
For applications that require polarization sensitivity the window should present low instrumental polarization. 

Thin Polypropylene, polyethylene, or thicker zotefoam have been used as vacuum window materials~\cite{Hanany2013,Bischoff_2013, 10.1117/12.672652}. 
The index of refraction of these materials is near 1.5 leading to relatively low reflective losses. 
Because of their 
low Young's modulus ($\lesssim 1~\mathrm{GPa}$~\cite{INEOS_PP_properties,INEOS_HDPE_properties}), the plastics-based materials bow under differential pressure, and their thicknesses are tuned to provide sufficient stiffness given the diameter. 
With fixed differential pressure, maximum window deflection is inversely proportional to Young's modulus, inversely proportional to the cube of the thickness, and proportional to the fourth power of the diameter~\cite{RoaksFormulas}.
Silicon and Quartz have also been used as window materials~\cite{Koller2022,Nagai:23} and with Young's moduli higher than polyethylene by a factor of more than $\sim$50 their bow can effectively be neglected. The index of refraction of silicon is 3.4 and to avoid reflectances exceeding 50\% an anti-reflectance coating (ARC) must be used.  
 
A common way to provide ARC on silicon optical elements is to fabricate subwavelength structures (SWS), and groups have used dicing~\cite{Datta:13,Golec:22}, chemical etching~\cite{Gallardo:17,Defrance:18,Macioce2020,Hasebe:21,Nagai:23}, and laser ablation~\cite{Schutz2016,matsumura2016,young_siliconSWS,matsumura2018,doi:10.1063/5.0022765,Takaku:21,wen2021,Kuroo:10,Brahm:14,ZHANG2016148,8419328,Sakurai:19,Yu:19,Koike:24} to make such SWS-ARC. Among these papers, only Nagai et al. report on a vacuum window operating at an astrophysical observatory~\cite{Nagai:23}. The authors made a silicon-based vacuum window for  the Atacama Submillimeter Telescope Experiment Band 10 receiver operating at 868.5~GHz with a fractional bandwidth of 19\%~\cite{Nagai:23}. They used chemical etching to make a two-layer sub-wavelength structures for an optically active diameter of 20~mm. The overall diameter was 40~mm.


In this paper we present the design, fabrication and characterization 
of a silicon-based vacuum window for the Deep Spectroscopic High-redshift Mapper (DESHIMA) 2.0 receiver on the ASTE telescope~\cite{10.1038/s41550-019-0850-8,10.1117/1.JATIS.5.3.035004,Taniguchi2022, 10.1117/1.JATIS.11.2.025007}. 
The ARC was based on laser-ablated SWS, a development that leverages our prior work with laser-ablated SWS with both silicon and other ceramics~\cite{Schutz2016,matsumura2016,young_siliconSWS,matsumura2018,doi:10.1063/5.0022765,Takaku:21,wen2021}. Sections~\ref{sec:req_and_design} and ~\ref{sec:fab} give the design, fabrication, and shape characterization. In Sections~\ref{sec:characterization} and~\ref{sec:results} we discuss the millimeter-wave transmission and reflection measurements and their results, and we summarize and conclude in Section~\ref{sec:conclusion}. To our knowledge, this paper is the first report of a fielded broadband silicon vacuum window for a millimeter-wave application.

\section{Requirements and Design}
\label{sec:req_and_design}


\subsection{Requirements}
\label{sec:req}

DESHIMA 2.0 is an instrument to observe astrophysical objects at the millimeter-submillimeter waveband.
The instrument consists of an on-chip spectrometer with a filter bank cooled to 120~mK and operating over a frequency band between 200 and 400~GHz.
The vacuum window of the cryogenic receiver is required to have a transmittance exceeding 90\% over the operating frequency band. 
The maximum incidence angle is about 4.25~degrees. This angle is small and therefore all transmittance and reflectance calculations are done at normal incidence. The physical window diameter is 124~mm and the optically active area has a diameter of 68~mm. The window must withstand multiple cycles to a pressure difference of up to 1~atm.  

\subsection{SWS-ARC Design}
\label{sec:sws_design}

\begin{figure}[h]
\begin{centering}
\includegraphics[width=1.0\linewidth]{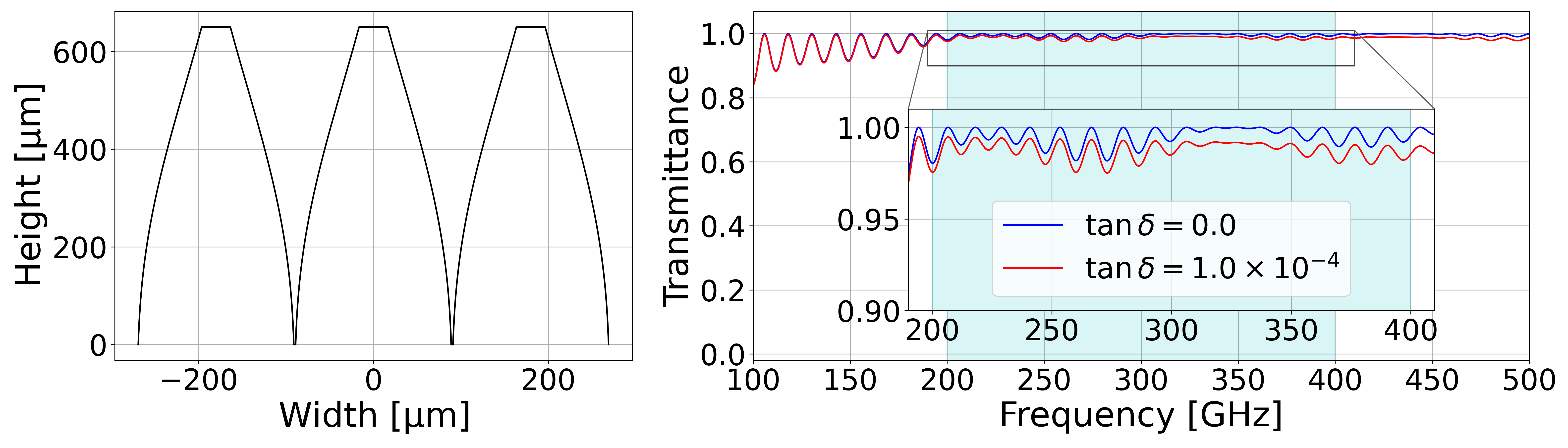}
\caption{The design shape of three elements in the periodic SWS (left) and the predicted transmittance of this ARC (right) with the DESHIMA pass-band highlighted (cyan). The vertical scale of the inset, between 0.9 and 1, highlights the $\sim$1\% difference between the spectrum without loss (blue), which gives an in-band average transmittance of 99.5\%, and the spectrum with loss (red), which gives an average of 98.8\%.   
\label{fig:swsdesign} }
\end{centering}
\end{figure}

High resistivity silicon has low loss in the millimeter-wave band. Most measurements give room temperature values $\tan \delta \lesssim 1\cdot 10^{-4}$~\cite{Lamb}. We used a boule of float zone silicon that had resistivity specification between 8.7~k$\Omega$-cm and 10.7~k$\Omega$-cm and for design purposes we assumed $\tan \delta = 1\cdot 10^{-4}$. 
To provide broad-band transmittance we designed a SWS-ARC with periodic, approximately pyramidal structure as shown in Figure~\ref{fig:swsdesign}. The height is 650~$\mu$m and the top width is 33~$\mu$m. 
We determined a pitch of $p = 180~\mathrm{\mu m}$ to avoid diffraction in the DESHIMA frequency band~\cite{Grann, Bruckner:07}. 
The shape of the SWS is based on an impedance matching model~\cite{Klopfenstein, Grann}, which is applied to minimize reflectances due to the transition between media with different indices of refraction. For the model we assumed $n=3.4$~\cite{Lamb} and $\Gamma = 0.05$ for the index of refraction of silicon and the shape parameter in the impedance matching model, respectively. The geometric shape shown in Figure~\ref{fig:swsdesign} has been derived from the index of refraction profile using second order effective medium theory~\cite{brawer}. 
To allow for SWS on both sides of the window, the thickness of the pre-ablated sample was 4.1 mm, a thickness we determined based on the expected in-band absorptance and the mechanical considerations outlined in Section~\ref{sec:MechDesign}. The substrate thickness of the ablated area is 2.8 mm and assuming a 50\% fill fraction for the SWS, the effective thickness is 3.45 mm. With that thickness we expect less than 1.1\% absorptance.

The predicted transmittance spectra without and with absorptive loss are given in Figure~\ref{fig:swsdesign}. The average transmittance without and with loss, are 99.5\% and 98.8\%, respectively, which give an indication of the efficacy of the SWS-ARC.

\subsection{Mechanical Design}
\label{sec:MechDesign}

\begin{figure}[h]
\centering
\includegraphics[width=0.6\linewidth]{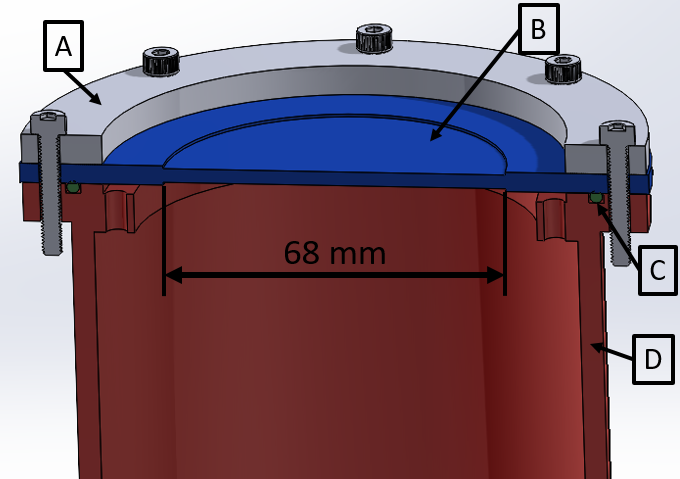}
\caption{Cross section of the vacuum window assembly. An aluminum ring (A, grey) clamps the 4.1~mm thick outer rim of the silicon window (B, blue) into a housing with an o-ring (D, red and C, green, respectively). The window's center area, with a diameter of 68~mm, has SWS-ARC. For that area, only the solid substrate of 2.8~mm thickness is shown.    \label{fig:mechdesign} } 
\end{figure}

The vacuum window was clamped with an aluminum ring into the dewar's aluminum entrance port with 8 bolts, and an o-ring provided the vacuum seal, see Figure~\ref{fig:mechdesign}.
We used a finite element analysis (FEA) \cite{ANSYSMechanical} 
to determine the window thickness. The FEA model included the overall mechanical configuration as shown in Figure~\ref{fig:mechdesign}, and fixed constraints were applied at the window housing. We compared the maximum equivalent Von-Mises stress to silicon's tensile strength as a function of window thickness.
For the purpose of FEA calculations we assumed that the area with SWS had only the unablated thickness and was thinner by 1.3~mm to allow for the SWS-ARC, see Section~\ref{sec:sws_design}.  
Four reported values for the tensile strength of silicon vary between 113 and 350 MPa~\cite{MindrumSilicon,UniversityWaferSilicon,AZOMProperties}.  
The chosen thickness of 4.1~mm gives a Von-Misses stress of 18~MPa, a factor of safety of more than 6 (9) relative to the minimum (median) reported value for  tensile strength of 113 (170)~MPa. The calculated maximum central deflection is 17~$\mu$m. We used a relatively large safety factor to hedge against the risk of the silicon cracking due to the ablation. 

\section{Sample Characterization}
\label{sec:fab}
\subsection{Sample preparation}

We cut and machined two slices from the silicon boule. One `flat' sample, with 100~mm diameter, was used for measurements of index and absorptive loss. The second, with 124~mm diameter, was patterned to be used as the vacuum window.
We measured the thickness of each sample in 20 locations and found an average and standard deviation of $t_{\mathrm{flat}} = 4.151\pm 0.005~\mathrm{mm}$ and $t_{\mathrm{window}} = 4.085\pm 0.004~\mathrm{mm}$, for the flat and the pre-machined vacuum window, respectively. 

\subsection{SWS}
\label{sec:fab_sws}

\begin{figure}[h]
\centering
\includegraphics[width=0.31\linewidth]{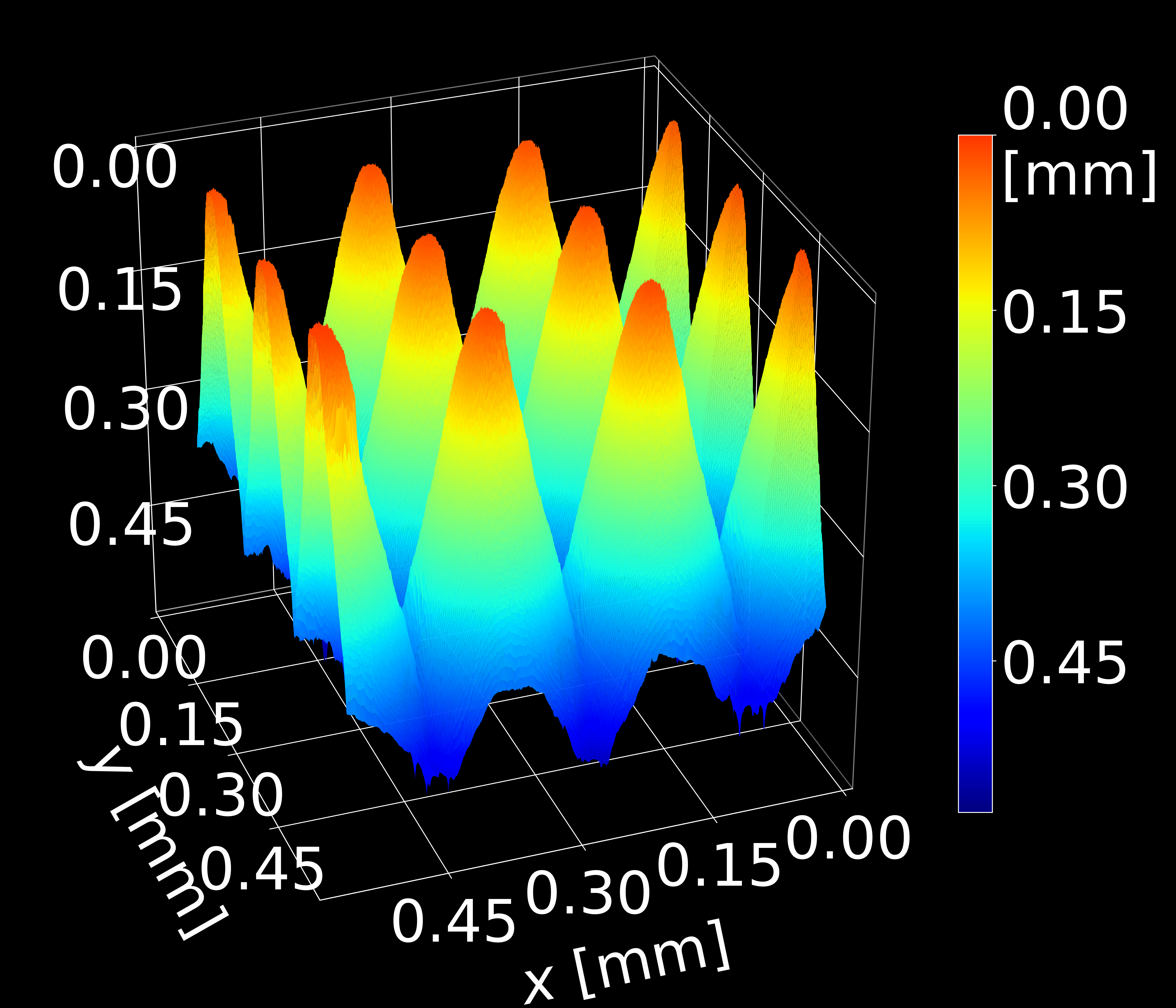}
\hspace{0.03in}
\includegraphics[width=0.315\linewidth]{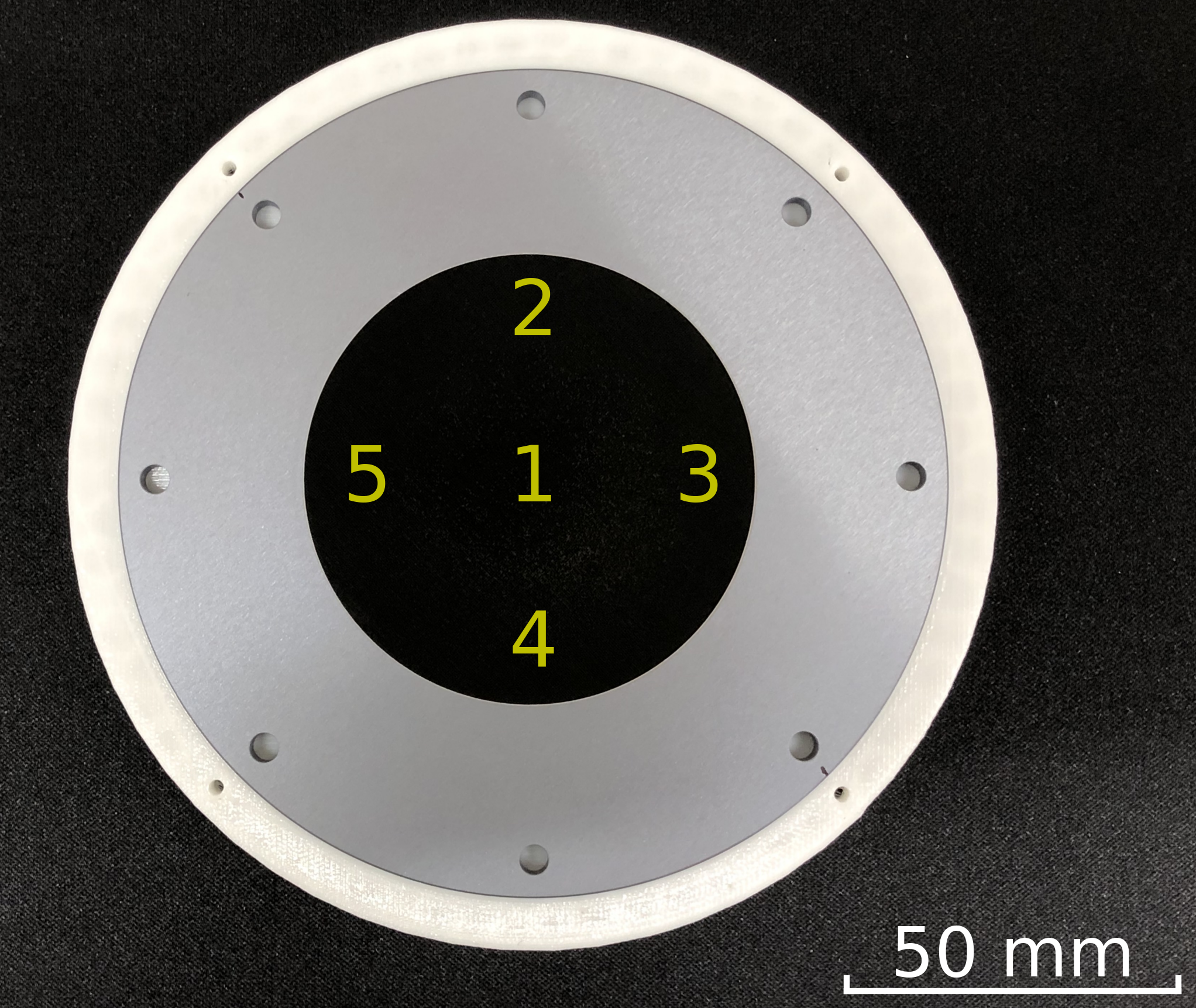}
\hspace{0.03in}
\includegraphics[width=0.315\linewidth]{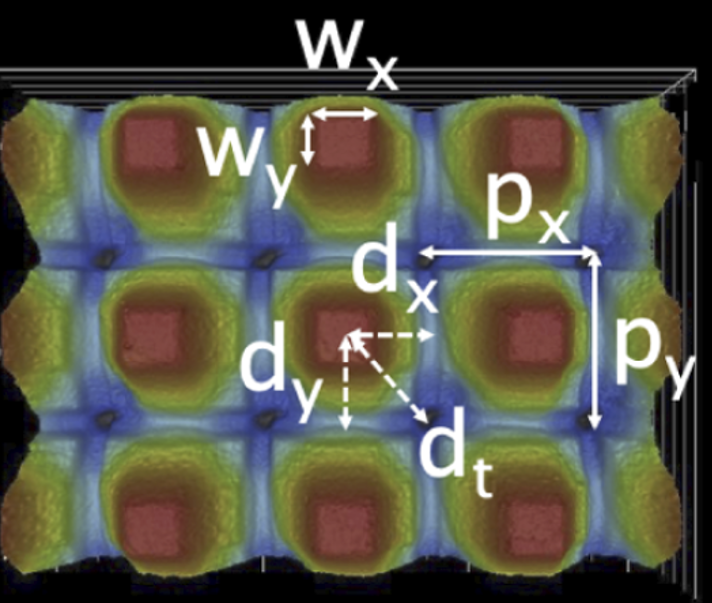}
\caption{A small section of the fabricated SWS-ARC (left), a photograph of the window (middle), and a sketch of the parameters used to characterize the SWS (right). In the middle panel, the black circle is the 68~mm diameter  optically active area with SWS and the silver ring around it is bare silicon. The numbers denote areas in which SWS shape measurements were conducted. In the right panel, ${\rm w}$ and ${\rm p}$ parameters are length measurements, and ${\rm d}$ are height measurements, see Table~\ref{tab:summaryfabshapes}.
}
\label{fig:swsfabricated}
\end{figure}

The SWS were ablated on both sides of the silicon disc using an ultrashort pulse laser. The laser model and some fabrication parameters are given Table~\ref{tab:laserpars}. 
The laser scan pattern was similar to patterns we used in the past for ablating silicon, alumina, and sapphire~\cite{karlJAP17,Takaku:21,doi:10.1063/5.0022765}.
To ensure more symmetric SWS-ARC pattern we rotated the sample by 90 degrees when we machined the second surface~\cite{doi:10.1063/5.0022765}.
A confocal microscope image of the fabricated structures and a photograph of the fully fabricated window are shown in  Figure~\ref{fig:swsfabricated}. 
We analyze the confocal microscope images to make measurements of SWS geometrical parameters at five approximately square areas on each side of the sample, each area with 63 pyramids.
The area locations and the definitions of the geometrical parameters are shown in Figure~\ref{fig:swsfabricated}. The measurements were binned to make histograms, the histograms were fit with Gaussians, and we give the Gaussian means $\mu$ and widths $\sigma$ in  Table~\ref{tab:summaryfabshapes} for the 315 pyramids per surface. The histograms are given in appendix~\ref{appendix_hist}. 

\begin{table}[h]
\centering
\caption{Parameters of the laser used for ablating the SWS-ARC.
\label{tab:laserpars}}
\renewcommand{\arraystretch}{1.0} 
\vspace{-0.25in}
    \begin{tabular}{l c}  \\
    \hline
    \multicolumn{2}{l}{Laser model: Carbide CB3 + CBM03-2H-3H} \\
    \hline
    Power (W) &  4 - 20\\
    Wavelength (nm) &  1030\\
    Pulse duration (fs) &  309$^{\dagger}$   \\
    GHz burst mode setting (sub-pulses)& 25$^{\dagger \dagger}$ \\
    Repetition rate (kHz) &   100 \\
    Beam spot diameter ($\mu$m) &  28$^{\dagger \dagger \dagger}$ \\ \hline
    \multicolumn{2}{l}{$^{\dagger}$ Burst mode.} \\
    \multicolumn{2}{l}{ $^{\dagger  \dagger}$ Proprietary technology of Light Conversion. } \\
    \multicolumn{2}{l}{ $^{\dagger  \dagger  \dagger}$ Diameter at $1/e^{2}$.}
    \end{tabular}
\end{table}

\begin{table}[h]
    \centering
    \caption{Measured Gaussian means $\mu$ and widths $\sigma$ for fits to histograms of the measured parameters $x_{i}$ of 315 pyramids per side. The fits are $f \propto \exp \left[ -(x_{i} - \mu)^{2}/2\sigma^{2} \right]$. The parameters $x_{i}$ are given in the right panel of Figure~\ref{fig:swsfabricated}. 
    \label{tab:summaryfabshapes} }
    \renewcommand{\arraystretch}{1.0} 
    \begin{tabular}{lc c}
    \hline
     \multirow{2}{*}{Parameter} & \multicolumn{2}{c}{$\mu\, \pm\, \sigma$ ($\mu$m)}  \\ 
     & Surface 1 & Surface 2 \\
     \hline
    Pitch in x (${\rm p_{x}}$) & 179 $\pm$ 2 & 179 $\pm$ 2 \\
    Pitch in y (${\rm p_{y}}$) & 179 $\pm$ 2 & 179 $\pm$ 2 \\
    Top width in x, (${\rm w_{x}}$) &  21 $\pm$ 2 & 22 $\pm$ 4  \\
    Top width in y, (${\rm w_{y}}$) & 19 $\pm$ 2 & 17 $\pm$ 2    \\
    Saddle height in x, (${\rm d_x}$) & 400 $\pm$ 9  & 400 $\pm$ 8 \\
    Saddle height in y, (${\rm d_y}$) & 401 $\pm$ 10  & 403 $\pm$ 8  \\
    Height, (${\rm d_t}$) & 581 $\pm$ 16 & 581 $\pm$ 11  \\
        \hline
    \end{tabular}
\end{table}




\section{Millimeter-wave measurement}
\label{sec:characterization}


\subsection{Optical measurement setup}

\begin{figure}
\centering
\includegraphics[width=0.5\linewidth]{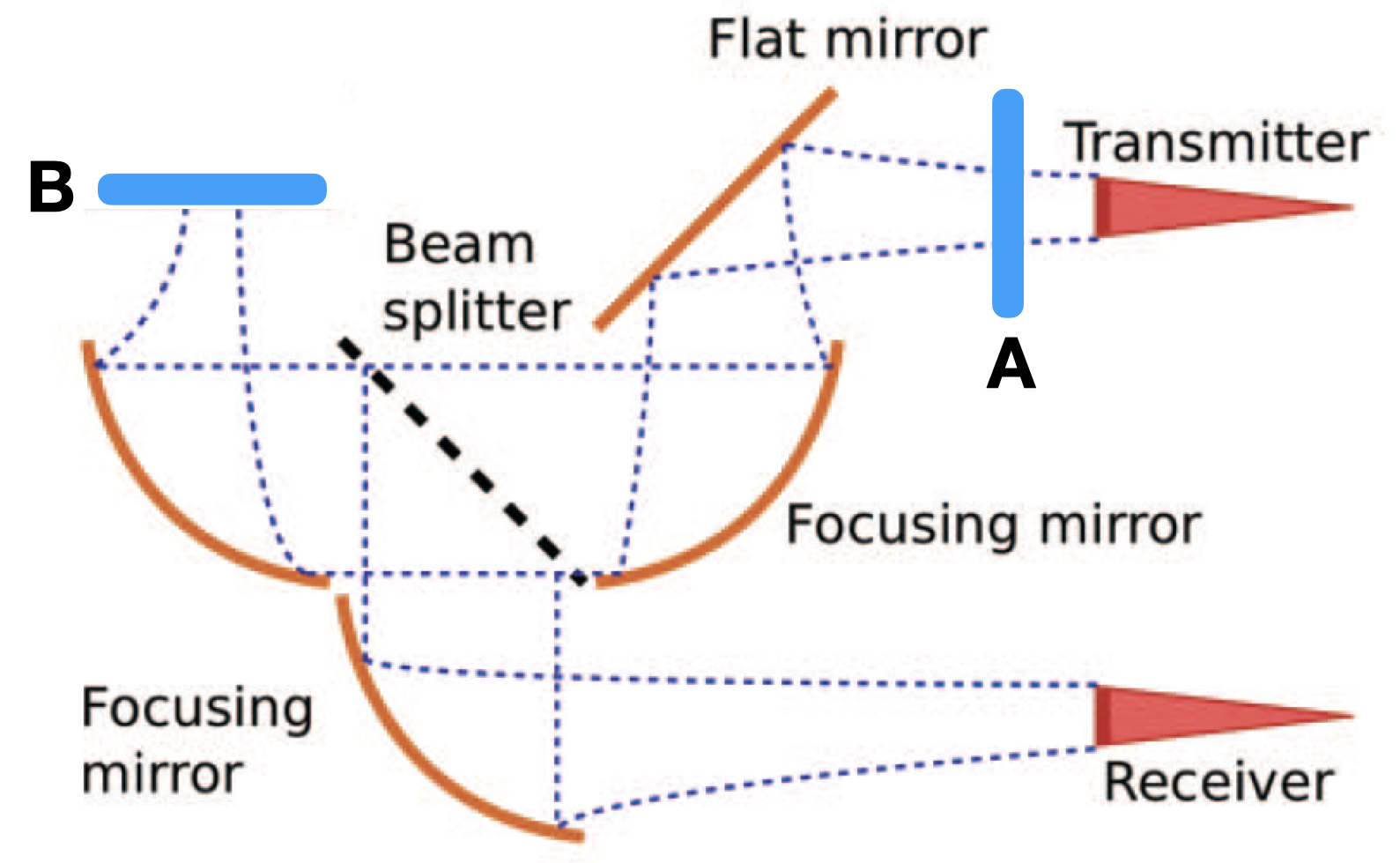}
\caption{Sketch of the experimental setup used for the reflectance and transmittance measurements. For transmittance: A is the sample and B is a flat mirror, and the measurement is normalized against a measurement without a sample. For reflectance: Nothing is placed in location A and B is the sample, and the measurement is normalized against a measurement when B is a mirror. \label{fig:experimentalsetup}}
\end{figure}

We measured the transmission and reflection of the flat and patterned plates using the measurement setup shown in Figure~\ref{fig:experimentalsetup}.
Measurements were conducted with a VNA in three frequency bands as listed in Table~\ref{tab:bestfit}. 
Transmission was measured as the ratio of signal when the sample was placed near the transmitter to the signal without a sample. reflection is the ratio of signal recorded with a sample to the signal recorded when the sample is replaced with a high reflectivity gold mirror, see Figure~\ref{fig:experimentalsetup}.To obtain transmittance $T$ and reflectance $R$ we square the signal ratios.  
All measurements were done at normal incidence. 
For the patterned sample we measured the transmittance and reflectance in two polarization states, parallel and perpendicular to the $x$ axis of one of the surfaces. 

\subsection{Data analysis}
\label{subsec:outlier_rejection}
We identify and reject data outliers by finding the best fit models to the $T$ and $R$ data with the index of refraction $n$ and loss $\tan\delta$ as fit parameters. We subtract the best fit model from the data, bin the residuals, fit them with Gaussians that have widths $\sigma_t$ and $\sigma_r$ for $T$ and $R$, respectively, and remove data points that were outside $3\sigma_{t,r}$. This process is repeated twice and we find that 2.4\% (14 points) , 0.8\% (6 points), and 8.7\% (103 points) of the data were removed per band from lowest to highest frequencies. 

We estimated the uncertainty in $n$ and $\tan \delta$ using a Monte-Carlo analysis. We used the standard deviation of the residual as a uniform uncertainty for all data points.  We generated 1000 mock datasets by adding Gaussian noise $\mathrm{N}(0,\sigma_t \, \mbox{or}\, \sigma_r)$ to each $T$ or $R$ data point. Each of the 1000 realizations was best fit to find 1000 values of $n$ and $\tan \delta$, we histogram the distributions, fit with Gaussians, and report their width $\sigma$ as the $1\sigma$ uncertainty per parameter.
We apply the same procedure to estimate the contribution of thickness measurement uncertainty. We fitted the data 1000 times, each time with a thickness drawn from a Gaussian distribution centered on the measured value and with width equal to the $1\sigma$ measurement uncertainty.
We apply the same procedure to estimate the contribution of thickness measurement uncertainty. We fitted the data 1000 times, each time assuming a thickness drawn from a Gaussian distribution centered on the measured value and with width equal to the $1\sigma$ measurement uncertainty. We obtained distributions for $n$ and $\tan \delta$ and we take the widths $\sigma$ of these distributions as the parameter uncertainties induced by a thickness uncertainty. We estimate the total parameter uncertainties by the quadrature sum of the statistical and thickness uncertainties.

When analyzing the transmission and reflection of the patterned sample we removed data points at the same frequencies that were flagged as outliers with the flat sample.

\section{Results}
\label{sec:results}

\subsection{Optical performance}

\subsubsection{Flat Sample}

\begin{figure}
\centering
\includegraphics[width=1.0\textwidth]{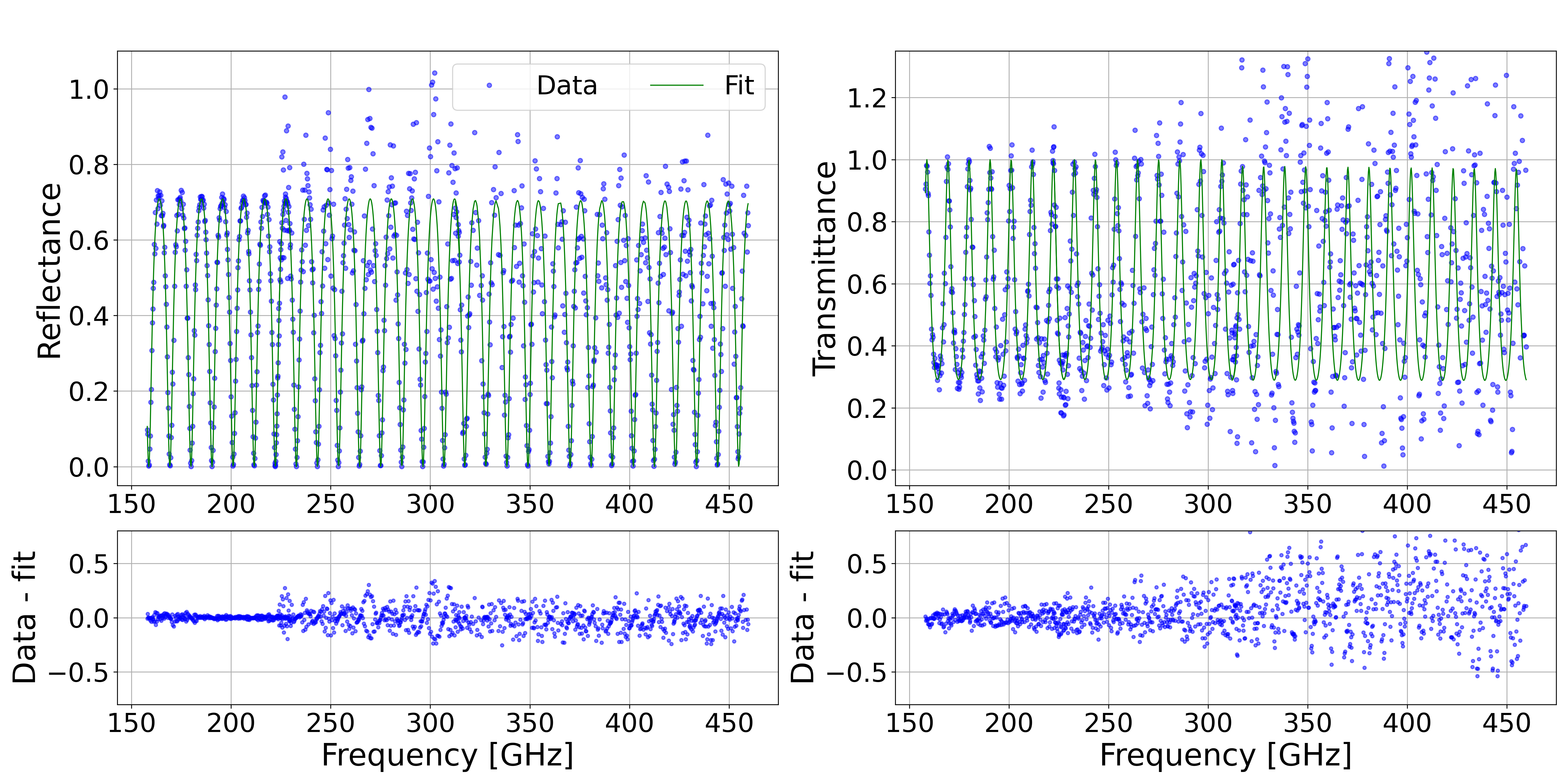}
\caption{Reflectance and transmittance spectra of the flat sample (data points, upper panels, left and right, respectively), the best fit model (line), and the residuals (lower panels). Values for the best fit $n$ and $\tan \delta$ are given in Table~\ref{tab:bestfit}. 
\label{fig:spectrum_pre}}
\end{figure}

The transmittance and reflectance spectra for the flat sample and the best fit model are shown in Figure~\ref{fig:spectrum_pre}. The best--fit and uncertainty values are summarized on Table~\ref{tab:bestfit}. 
The best-fit indices of refraction are consistent with each other to within $1\sigma$ over the entire frequency range. The loss tangents are consistent with zero and with each other within $1.5\sigma$. The uncertainties at the two lower frequency band suggest a loss smaller than $1\times10^{-4}$ at a statistically significant level.


\begin{table}
    \centering
    \caption{The best--fit of the refractive index $n$ and loss $\tan\delta$ for each frequency band.\label{tab:bestfit} }
    \renewcommand{\arraystretch}{1.0}
        \begin{tabular}{c | c c | c c }
        \hline
        Frequency band [GHz] & \multicolumn{2}{c|}{Index $n$} & \multicolumn{2}{c}{Loss $ \tan\delta ~(\times 10^{4})$} \\\hline
                          & fit              & RCWA input & fit & RCWA input\\\hline
        158 -- 230 & $3.412\pm 0.004$ &  3.412     & $0.0\pm 0.2$     & 0 \\
        222 -- 315 & $3.412\pm 0.002$ &  3.412     & $0.0\pm 0.3$ & 0 \\
        312 -- 460 & $3.414\pm 0.002$ &  3.414     & $1.2\pm 0.8$   & 1.2 \\\hline
    \end{tabular}
\end{table}

\subsubsection{Patterned Sample}

\begin{figure}
\centering 
\includegraphics[width=1.0\linewidth]{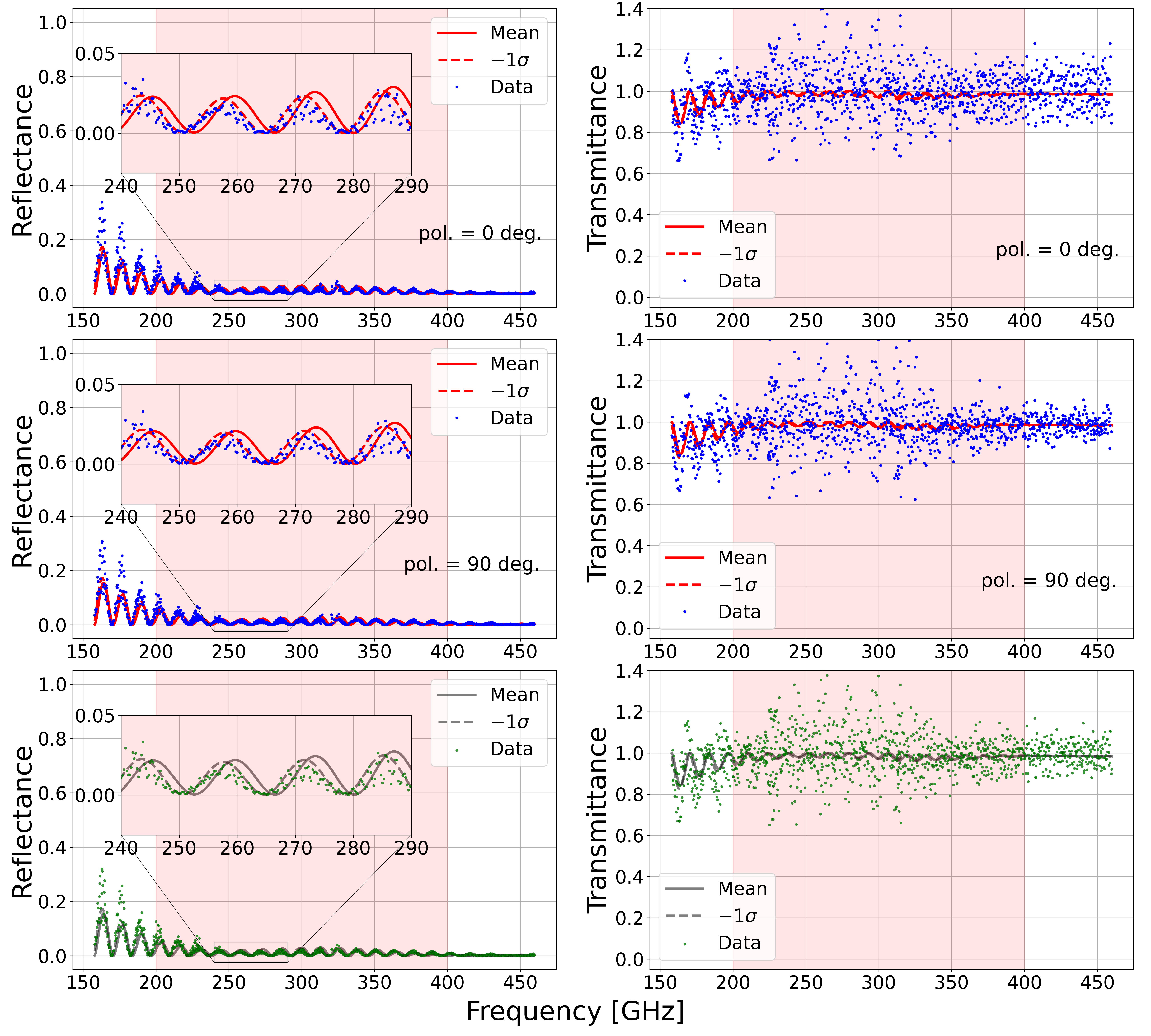}
\caption{Reflectance (left panels) and transmittance (right panels) spectra (dots) of the patterned sample for polarization angles of 0 (top panels) and 90 degrees (middle panels), and the average of the two polarizations (bottom panels). Lines give the results of RCWA simulations assuming the average pyramid using the mean values for the shape parameters given in Table~\ref{tab:summaryfabshapes} (solid, labeled mean), and with the heights ${\rm d}$ shorter by $1\sigma$ (dash, labeled $-1\sigma$). The insets on the left highlight the small $\lesssim 2.5\%$ reflectance oscillations and the general agreement with the RCWA simulation.
\label{fig:spectrum_post}
}
\end{figure}

Fig.~\ref{fig:spectrum_post} shows the reflectance and transmittance of the patterned sample in two polarization states and after we average the two polarizations.
Referring to the average of the two polarizations, average in-band 200--400~GHz transmittance is 99.6\% and the average reflectance is 1.3\%. The maximum reflectance is below 10\% and the maximum modeled reflectance, which we describe below, is 6\%. The averaged transmittance for polarization angles of 0 and 90~degrees are 100.0\% and 99.1\%, respectively.
Absorptive loss is undetectable.

We compare the measurements to predictions using rigorous coupled-wave analysis (RCWA)~\cite{Moharam:95}. We use measurements from 63 pyramids to form a 3-dimensional solid model of an `average pyramid'. 
Two versions of this model are inserted into an RCWA software~\cite{rsoft_diffractmod}, one that has the mean values listed in Table~\ref{tab:summaryfabshapes} and a second with identical ${\rm p}$ and ${\rm w}$ parameters, but the heights ${\rm d}$ are reduced by their $1\sigma$ dispersion. Using periodic boundary conditions and the $n$ and $\tan \delta$ given in Table~\ref{tab:bestfit} we calculate the expected transmittance and reflectance spectra. 
The spectra are included in Fig.~\ref{fig:spectrum_post} and they are largely consistent with the data. The reflectance data show a 1-2~GHz phase shift with the `mean' model, but good agreement with the `$-1\sigma$' model.  Both RCWA models give reflectance amplitudes that are consistent with the measured data.

\begin{figure}
    \centering
    \includegraphics[width=1.0\linewidth]{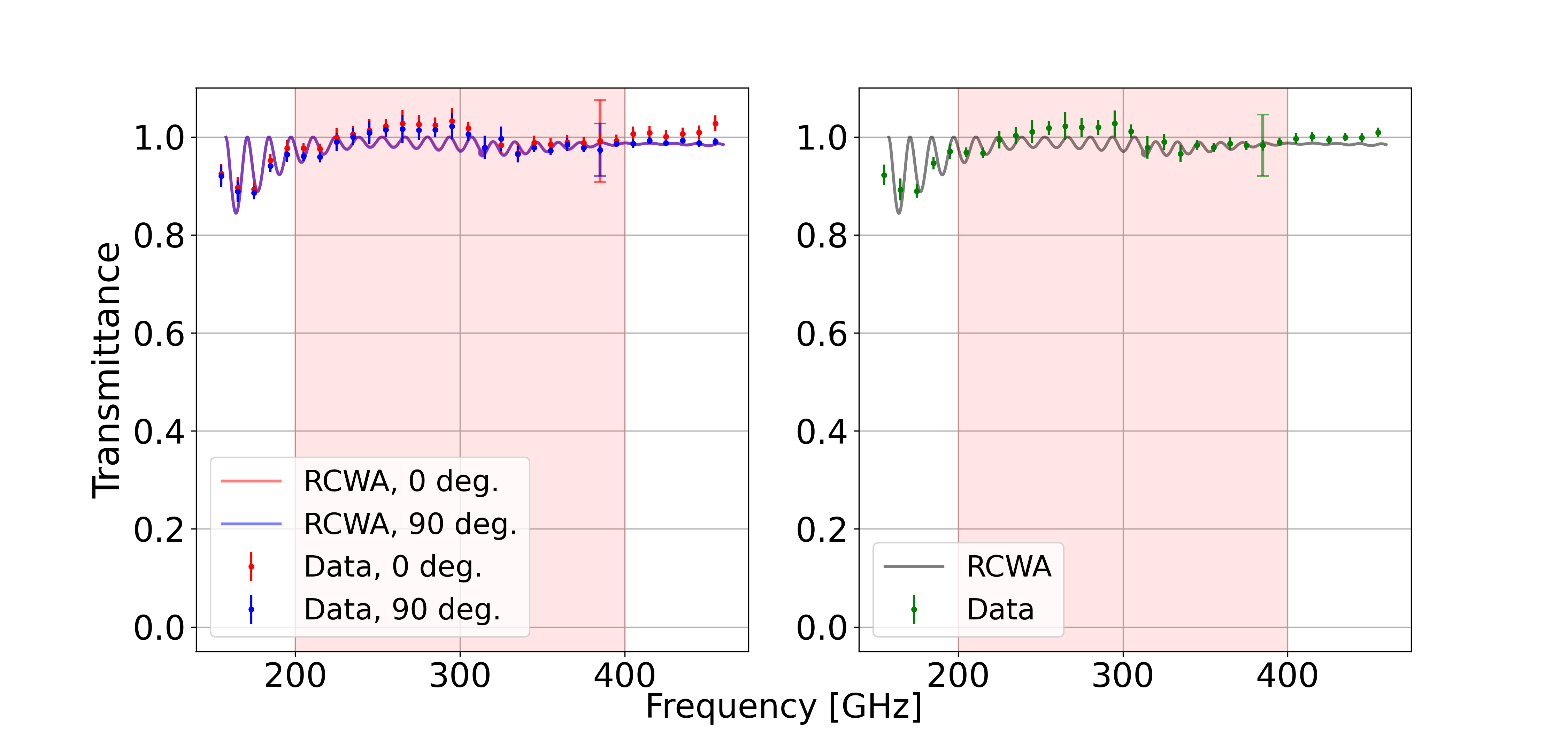}
    \caption{Transmittance spectra of the patterned sample binned into 10~GHz bins for polarization angles 0 and 90 degrees (left), their average (right), and RCWA predictions (solid). Smaller error bars indicate the uncertainties on the mean in each bin. A single larger error bar per polarization state per panel indicates a typical $1\sigma$ data dispersion per bin. }
    \label{fig:trans_bin}
\end{figure}

To reduce noise in the transmission spectra we bin the data and compare them to a similarly binned RCWA model. The frequency bins are 10~GHz wide and in Fig.~\ref{fig:trans_bin} we display the bin averages, the errors on the mean, and one  standard deviation to indicate the dispersion of the binned data.

\subsection{Mechanical Robustness}

The mechanical robustness of the window has been tested by placing the window within its receiver mount and subjecting it to a differential pressure of 1~atm. No attempt was made to regulate the pumping speed and a differential pressure of essentially 1~atm was achieved in a few minutes. The cycle has been repeated four times. A helium leak test did not find leaks. 

After the window was delivered, it has been pumped in the lab and while integrated into the ASTE telescope about half-dozen times. The DESHIMA receiver operated under vacuum for about a year. There were no vacuum leaks nor other incidents throughout these operations.

\section{Discussion and Conclusions }
\label{sec:conclusion}

\begin{figure}
\centering
\includegraphics[width = 0.7\textwidth]{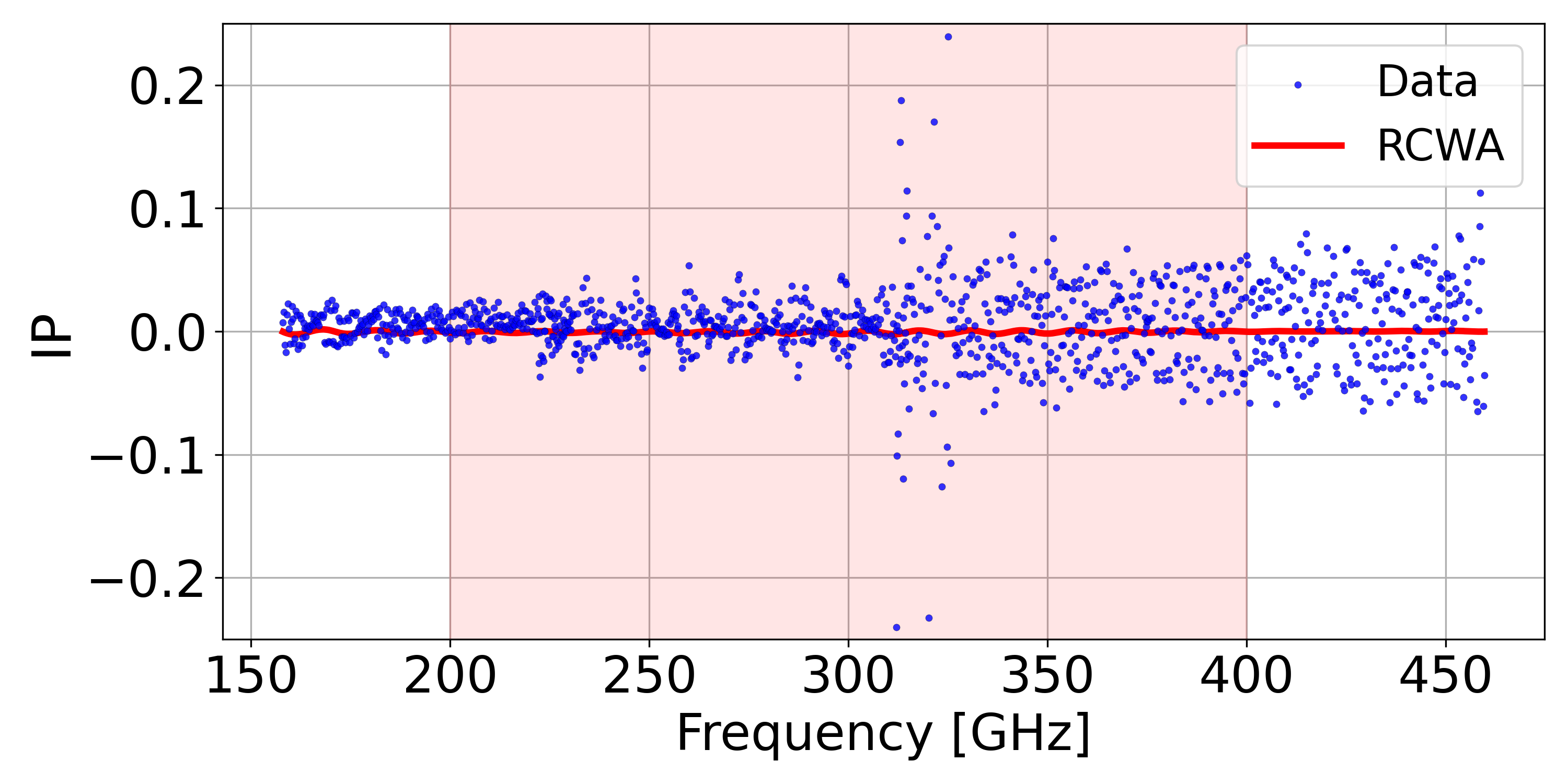}
\caption{Instrumental polarization (IP) of the patterned sample\label{fig:ip}.The outlier rejection is applied before calculating IP as described in Section \ref{subsec:outlier_rejection}. }
\end{figure}

Although the DESHIMA science goals don't require polarization sensitivity, other astrophysical measurements in this frequency band may require it, notably measurements of the cosmic microwave background or of Galactic dust~\cite{Taurus:2024dyi,10.1117/12.2312985}. As described in Section~\ref{sec:fab_sws}, we rotated the sample by 90~degrees when fabricating the second side to symmetrize possible $x,y$ asymmetry in the ablation process. Such asymmetries can give differential reflectance between the two polarization states resulting in instrumental polarization. We quantify the instrumental polarization spectrum ${\rm IP}(\nu)$ through
\begin{align}
    \mathrm{IP}(\nu) = \frac{T(\nu)|_{\mathrm{p=0}} - T(\nu)|_{\mathrm{p=90}}}{T(\nu)|_{\mathrm{p=0}} + T(\nu)|_{\mathrm{p=90}}},
\end{align}
where $T$ is the measured transmittance and ${\rm p}=0,\, 90$ indicate polarization states. The results are given in Fig.~\ref{fig:ip}. The noisy data give weak constraints, but the RCWA model gives an average ${\rm IP} = 0.05\%$ and a maximum ${\rm IP} < 0.09\%$ over the DESHIMA band.

We made a 68~mm optically-active diameter,  millimeter-wave transparent, vacuum window made of high resistivity silicon. The window was designed to operate with 67\% bandwidth centered on 300~GHz, which is a wavelength of 1~mm. To reduce in band reflectance values exceeding 60\% we fabricated SWS-ARC using laser ablation. We measured an average in band transmittance and reflectance of 99\% and 1\%, respectively. The highest reflectance in band is less than 10\% satisfying the design requirement. We constructed an RCWA model based on the measured shape of the SWS and the model is in agreement with the measured data. 
The vacuum window was implemented with the DESHIMA receiver, which conducted astrophysical observations for over a year. To our knowledge, this is the first report of a silicon vacuum window with laser-ablated SWS fielded with a millimeter-wave astrophysics telescope. Implementation of the this technology for frequencies of few THz is relatively straightforward because SWS have already been demonstrated at 3~THz~\cite{Kuroo:10}. Fabrication of SWS at lower frequencies has also been demonstrated with various techniques~\cite{Datta:13,Gallardo:17,Defrance:18,Macioce2020,Hasebe:21,Nagai:23,young_siliconSWS}, however, high resistivity silicon is only readily commercially available at diameters $\lesssim 20~\mathrm{cm}$ currently limiting the size of silicon-based vacuum windows.  

\appendix

\section{Detailed Structure measurements}
\label{appendix_hist}

\begin{figure}[h]
\centering
\includegraphics[width = 1.0\linewidth]{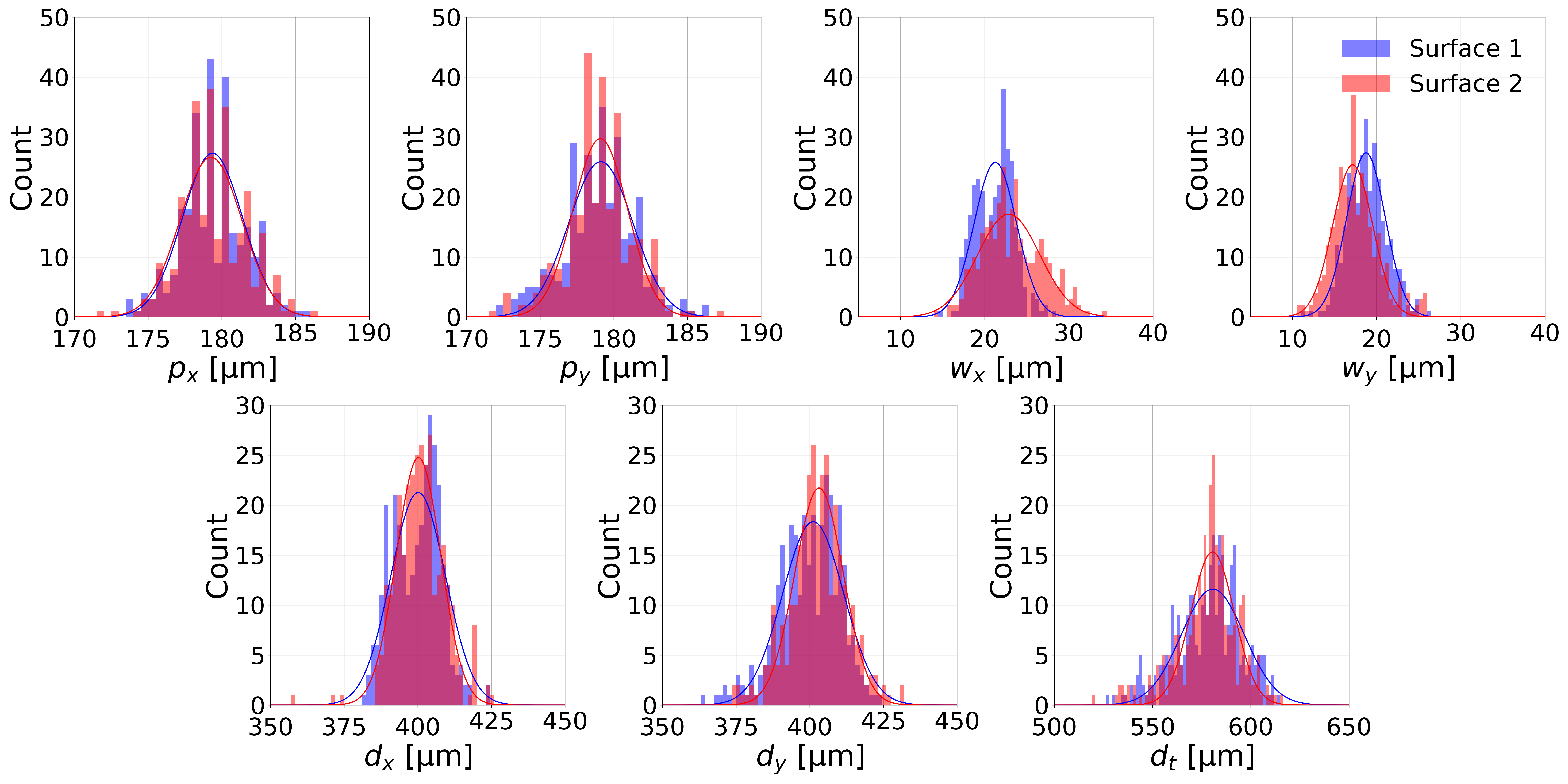}
\caption{Histograms of the measured parameters for 315 pyramids for each of surface 1~(blue) and 2~(red). The solid curves show Gaussians with means and widths, which are reported in  
Table~\ref{tab:summaryfabshapes}.
\label{fig:geometry_histogram} }
\end{figure}

In Section~\ref{sec:fab_sws} and Table~\ref{tab:summaryfabshapes} we report the average values of the SWS structure parameters derived for all 315 pyramids per patterned sample side. The uncertainties reported in the Table are the standard deviation of Gaussian fits to histograms of the shape parameters. The histograms are given in   Figure~\ref{fig:geometry_histogram}.

\bibliography{bibtex}

\end{document}